\documentclass[usenatbib]{mn2e}
\input{psfig}
\usepackage{natbib} 
\usepackage{varioref}

\begin{document} 
\def\lsim{\mathrel{\hbox{\rlap{\hbox{\lower4pt\hbox{$\sim$}}}\hbox{$<$}}}}
\def\gsim{\mathrel{\hbox{\rlap{\hbox{\lower4pt\hbox{$\sim$}}}\hbox{$>$}}}}
\def\simlt{\mathrel{\rlap{\lower 3pt\hbox{$\sim$}}
        \raise 2.0pt\hbox{$<$}}}
\def\simgt{\mathrel{\rlap{\lower 3pt\hbox{$\sim$}}
        \raise 2.0pt\hbox{$>$}}}

\title[The $M_\bullet$ - $T$ equilibrium relation]
{The $M_\bullet$ - $T$ relation for a black hole in thermodynamic equilibrium 
with the surrounding intergalactic medium}
\author[A. Cattaneo]
{A.~Cattaneo $^{1}$\\
\\
$^1$Astrophysikalisches Institut Potsdam, an der Sternwarte 16, 14482 Potsdam, Germany\\}

\maketitle 
\begin{abstract}
I consider a toy model of self-regulated black hole accretion.
The black hole grows through Bondi accretion and a fraction of the accretion power is 
distributed as thermal feedback into the surrounding gas.
The gas expands or contracts until AGN heating and radiative cooling balance
each other.
The balance of heating and cooling is used to determine
a quasi-equilibrium temperature at which the black hole accretes in self-regulated equilibrium with the
surrounding intergalactic medium. 
This temperature grows with the black hole mass.
The temperature increase is very steep around a critical black hole mass due to the shape of the cooling function. 
The quasi-equilibrium temperature
cannot exceed the virial temperature or the AGN will drive a thermal wind.
This limits the black hole mass to a maximum value determined by the depth of the potential well.
In the regime in which cooling is dominated by  bremsstrahlung,  this model determines a relation between
black hole mass and halo characteristic velocity of the form $M_\bullet\propto v^4$.
The predictions of the model are consistent with the observed black hole mass -- bulge velocity dispersion relation.
\end{abstract}

\newpage
\section{Introduction}

All massive spheroids for which the data are sufficiently good show evidence for a central
black hole. The relation between black hole mass and bulge mass 
\citep{marconi_hunt03,haering_rix04}
suggests that the black hole contains $\sim 0.2\%$ of the bulge stellar mass.
The gravitational binding energy released by matter accreted onto the black hole is 
$\sim 10\%$ of the rest mass energy. A small fraction of this energy is sufficient to affect
the dynamics and the thermal state of the intergalactic medium (IGM) dramatically.

Active galactic nucleus (AGN) heating was introduced to solve the cooling flow problem
in galaxy clusters
 \citep{tabor_binney93,binney_tabor95,tucker_david97,ciotti_ostriker97,
cavaliere_etal02,
ruszkowski_begelman02} 
and the related problem of the entropy floor
\citep{roychowdhury_etal04,lapi_etal05}.
\citet{silk_rees98} went further and proposed that AGN outflows are an essential 
part of the galaxy formation process.
In this scenario, it is AGN feedback that determines the observed relation between black hole mass
and velocity dispersion of the host bulge \citep{ferrarese_merritt00,gebhardt_etal00,merritt_ferrarese01,tremaine_etal02}.
\citet{king03}, \citet{granato_etal04}, \citet{murray_etal05} and \citet{dimatteo_etal05}
have developed more elaborate versions of the original idea by \citet{silk_rees98} that AGN winds terminate star formation
in the host galaxy (the reference list does not pretend to be complete).

The interaction of the black hole with the IGM is a complex problem, which
requires hydrodynamic simulations
(e.g. \citealp{quilis_etal01, reynolds_etal01,churazov_etal02,reynolds_etal02,
basson_alexander03,omma_binney04,ruszkowski_etal04,
brueggen_etal05,dimatteo_etal05,zanni_etal05}),
but here I want to concentrate on a simple aspect,
where an analytic approach is possible and can offer a better insight into the results
of hydrodynamic simulations.

I consider a black hole at the centre of a profile with a core,
where the gas is approximately homogeneous and isothermal.  
The black hole accrete mass and returns heat to its surrounding environment.
The core expands or contracts until AGN heating and radiative cooling balance each other.
The balance of heating and cooling determines the relation between
the mass of the black hole and the temperature of the IGM at the equilibrium.
If the equilibrium temperature is much larger than the virial temperature, the system will
only admit outflow solutions.

\section{Model of the BH -- IGM coupling}

A black hole of mass $M_\bullet$ 
surrounded by gas with density $\rho$ and speed of sound $c_{\rm s}$ accretes at a
rate determined by the \citet{bondi52} formula:
\begin{equation}
\label{bondi}
\dot{M}_\bullet=4\pi\alpha({\rm G}M_\bullet)^2\,{\rho\over c_{\rm s}^{3}}=
4\pi\alpha({\rm G}M_\bullet)^2\,\rho\left(\gamma{kT\over\mu m_{\rm p}}\right)^{-{3\over 2}},
\end{equation}
where G is the gravitational constant and $\alpha$ is a number of order unity.
In the second equality, the speed of sound $c_{\rm s}$ is rewritten in terms of the temperature $T$.
Here $\gamma$ is the adiabatic index, $k$ is the Boltzmann constant, $m_{\rm p}$ is the
proton mass and $\mu$ is the mean particle mass as a fraction of the proton mass.
Eq.~(\ref{bondi}) assumes that the accretion is spherical
and that the gas is approximately static at infinity, but  is widely used to estimate $\dot{M}_\bullet$ 
even when these conditions are not satisfied (e.g. \citealp{dimatteo_etal05}).

A fraction $\beta$ of the power generated by the accretion of matter onto the black hole heats the 
surrounding gas 
at a rate $\dot{Q}_{\rm heat}$. If the heat is
distributed over a gas mass $M_{\rm gas}$, then the heating rate per unit mass is:
$${\dot{Q}_{\rm heat}\over M_{\rm gas}}=
{\beta\epsilon\dot{M}_\bullet{\rm c}^2\over M_{\rm gas}}=
4\pi\alpha\beta\epsilon{\rm c}^2\,
({\rm G}M_\bullet)^2\,{\rho\over M_{\rm gas}}\left(\gamma{kT\over\mu m_{\rm p}}\right)^{-{3\over 2}}=$$
\begin{equation}
\label{heating}
=630\,\alpha\beta\,\epsilon_{0.1}\gamma_{5\over 3}\mu_{16\over 27}X_{3\over4}^{-1}\,
{M_{\bullet\,8}^2\over M_{\rm gas\,8}}\,
n_{\rm H}T_6^{-3/2}{\rm\,erg\,s^{-1}cm^3g^{-1}},
\end{equation}
where $c$ is the speed of light and $\epsilon\sim 0.1$ is the energetic efficiency of black hole accretion.
In the second line, where we insert the numerical values,
$\epsilon_{0.1}\equiv\epsilon/0.1$,
$\gamma_{5/3}\equiv3\gamma/5$,
$\mu_{16/27}\equiv27\mu/16$,
$X_{3/4}\equiv4X_{\rm p}/3$ ($X_{\rm p}$ is the hydrogen baryon fraction),
$M_{\bullet\,8}\equiv M_\bullet/10^8M_\odot$,
$M_{\rm gas\,8}\equiv M_{\rm gas}/10^8M_\odot$ and
$T_6\equiv T/10^6{\rm K}$, while
$n_{\rm H}$ is the number density of hydrogen atoms (free protons for an ionised gas).

The cooling rate per unit mass of the heated gas is:
\begin{equation}
\label{cooling}
{\dot{Q}_{\rm cool}\over M_{\rm gas}}={1\over\rho}\,{\Lambda}(T,Z)\,n_{\rm H}^2=
{X_{\rm p}\over m_{\rm p}}\,{\Lambda}(T,Z)\,n_{\rm H}= 
\end{equation}
$$=45\,X_{3\over4}\,\Lambda_{-22}(T,Z)\,n_{\rm H}{\rm\,erg\,s^{-1}cm^3g^{-1}},$$
where $\Lambda n_{\rm H}^2$ is the radiated power per unit volume.
$\Lambda$ is a function of the temperature $T$ and the metallicity $Z$, and
$\Lambda_{-22}\equiv\Lambda/10^{-22}{\rm\,erg\,s^{-1}cm^3}$.
In the next Section I proceed to examine under what conditions the heating term and the cooling term balance each other off.

\section{Local $M_\bullet$ -- $T$ equilibrium relation}

Both the heating (Eq.~\ref{heating}) and the cooling (Eq.~\ref{cooling}) rate scale linearly with the 
density of the gas. The difference is in the temperature scaling. 
Fig.~1 compares the the rates of heating and cooling as a function of temperature.
The curves are computed from the collisional ionisation equilibrium cooling function of \citet{sutherland_dopita93} and show $\dot{Q}_{\rm cool}(T)/(n_{\rm H}M_{\rm gas})$ for different 
metallicities.
The diagonal straight lines show the heating term $\dot{Q}_{\rm heat}(T)/(n_{\rm H}M_{\rm gas})$
and are parameterised by the black hole -- IGM coupling factor
$\zeta=\alpha\beta\delta{M_{\bullet\,8}^2/M_{\rm gas\,8}}$ (Eq.~\ref{heating}),
where $\delta$ is the ratio between the gas density at the Bondi radius (in Eq.~\ref{heating}) and the gas density at the radius
that contains a gas mass $M_{\rm gas}$ (in Eq.~\ref{cooling}). 

\begin{figure}
\centerline{\hbox{
\psfig{figure=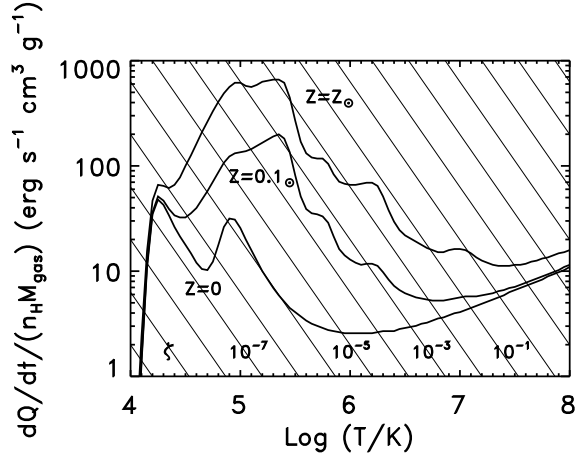,height=6cm,angle=0}}}
\caption{$\dot{Q}_{\rm heat}(T,\zeta)/(n_{\rm H}M_{\rm gas})$ (diagonal lines) and
$\dot{Q}_{\rm cool}(T,Z)/(n_{\rm H}M_{\rm gas})$ (curves) for different values of the black hole -- IGM
coupling constant $\zeta$ and the metallicity $Z$.}
\end{figure}

At very low $T$, the heating term always dominates and the temperature increases.
At very high $T$, the cooling term dominates and the temperature goes down.
In both cases, $T$ will converge to an equilibrium temperature $T_{\rm eq}$ determined by two parameters:
$Z$ and $\zeta$. 

For each metallicity, there is a value of $\zeta$, which we call $\zeta(Z)$, where the equilbrium 
extends over a temperature range (Fig.~1). For zero metals, 
this value is $\zeta(Z)\simeq 10^{-3}$ and the temperature range is $10^5{\rm\,K}\lsim T_{\rm eq}\lsim 3\times 10^5{\rm \,K}$. For $Z=0.1\,Z_\odot$, 
$\zeta(Z)\simeq  0.025$ and the temperature range is $2\times 10^5{\rm\,K}\lsim T_{\rm eq}\lsim 10^6{\rm\,K}$.
For Solar abundances,
$\zeta(Z)\sim 0.1$ and the temperature range is $2\times10^5{\rm\,K}\lsim T_{\rm eq}\lsim 3\times 10^6{\rm\,K}$.
If the parameter $\zeta$, which measures the effective efficiency of AGN heating, is
$\zeta<\zeta(Z)$, then cooling dominates at all but the lowest temperatures and AGN feedback is negligible.

For $\zeta\gg\zeta(Z)$, cooling is dominated by bremsstrahlung. 
In this regime, the cooling function can be approximated as $\Lambda_{-22}\sim 0.2\sqrt T_6$,
so that equating (\ref{heating}) and (\ref{cooling}) gives:
\begin{equation}
\label{teq}
T_{\rm eq}\sim 8\times 10^6\,(\alpha\beta\delta\,\epsilon_{0.1}\gamma_{5\over 3}\mu_{16\over 27})^{1\over 2}X_{3\over4}^{-1}\,
{M_{\bullet\,8}\over M_{\rm gas\,8}^{1\over 2}}{\rm\,K}.
\end{equation}
The equilibrium temperature is proportional to the black hole mass. The constant of proportionality depends on
the black hole accretion efficiency with respect to the Bondi rate, on the energetic efficiency of AGN heating and on the gas mass on which the feedback is distributed.

The equilibrium described in this Section is a local quasi-equilibrium. 
It is a quasi-equilibrium because $T_{\rm eq}$ is not constant with time.
The black hole accretes at the rate specified by Eq.~(\ref{bondi}) 
and $T_{\rm eq}$ must constantly increase to keep up with the increasing black hole mass ($\zeta\propto M_\bullet^2$). 
However, substituting Eq.~(\ref{teq}) into Eq.~(\ref{bondi}) shows that the
quasi-equilibrium accretion rate is $\dot{M}_\bullet\propto M_\bullet^{1/2}$, while the accretion rate without
self-regulation would be $\dot{M}_\bullet\propto M_\bullet^{2}$ until the Eddington limit is reached.
This model predicts that in the self-regulated regime $\dot{M}_\bullet/\dot{M}_{\rm Eddington}\propto M_\bullet^{-1/2}$.
The equilibrium is local because we made no assumption for the scale on which the feedback is distributed,
which may be as small as the nuclear region or as large as a galaxy cluster.
In the next Section we shall see how relating the feedback scale to the properties of the host system can generate a limit mass for black hole growth.

\section{The $M_\bullet$ -- $T_{\rm vir}$ equilibrium relation in galaxy clusters}

If AGN heating is relevant not only for the self-regulation of the black hole and for the gas in the central $\sim 1\,$kpc, 
but also to keep the cluster gas hot, then the feedback must be distributed on a scale of the order of the radius of the cluster core ($\sim 100\,$kpc).
In this scenario, one assumes that $T_{\rm eq}\sim T_{\rm vir}$ and that $M_{\rm gas}$ is equal to the mass of the gas in the core of the hot gas distribution, and one uses the equation
$\dot{Q}_{\rm heat}(T)/(n_{\rm H}M_{\rm gas})=\dot{Q}_{\rm cool}(T)/(n_{\rm H}M_{\rm gas})$
to determine the equilibrium black hole mass.

\citet{komatsu_seljak01} have assumed that the dark matter distribution in a virialised halo is described
by the \citet{navarro_etal97} profile and have used this profile to compute analytic hydrostatic 
equilibrium solutions for the baryonic component
(I verified with computer simulations that these are truly hydrostatic solutions). They contain three parameters: the virial mass $M_{\rm vir}$, 
the virial radius $r_{\rm vir}$ and the mass fraction in hot gas, which, for a cluster, is of the order of the cosmic baryon fraction ($\sim 0.1$).
The first two are not independent when a critical density contrast 
(a redshift of collapse) is specified.
The relation between the mass of the black hole $M_\bullet$ and the virial velocity $v_{\rm vir}={\rm G}M_{\rm vir}/r_{\rm vir}$
is determined from Eq.~(\ref{teq}), where
$T_{\rm eq}\sim{\rm G}\mu m_{\rm p}M_{\rm vir}/(3kr_{\rm vir})$ and 
$M_{\rm gas}$ is determined by the \citet{komatsu_seljak01} model.

In an isothermal sphere the virial velocity $v_{\rm vir}$ and the velocity dispersion $\sigma$ are related by $v_{\rm vir}=\sigma\sqrt 2$.
Fig.~2 shows the $M_\bullet$ -- $\sigma$ relation between the mass of the black and the velocity dispersion of the host galaxy
if the stars in the bulge of the central galaxy have the same velocity dispersion as the dark matter.
The two lines correspond to different redshifts of collapse of the dark matter halo.
The redshift of collapse $z_{\rm c}$ is used to calculate the virial overdensity
with the fitting formulae of \citet{bryan_norman98} and thus to compute $r_{\rm vir}$ as a function of $M_{\rm vir}$.
The data points show the observed black hole mass -- bulge velocity dispersion relation.

The model and the data overlap already in the simple case $\alpha\sim\beta\sim\delta\sim 1$ (Fig.~2). One must exert great caution in relating the virial velocity of the cluster $v_{\rm vir}$ to the velocity dispersion of the central galaxy $\sigma$. A more careful analysis would show that 
$v_{\rm vir}/\sqrt 2$ is an overestimate of the velocity dispersion of the central galaxy, which is compensated by an underestimate of $\delta$, since the gas density is higher at the centre due to the presence of the central galaxy. 

\begin{figure}
\centerline{\hbox{
\psfig{figure=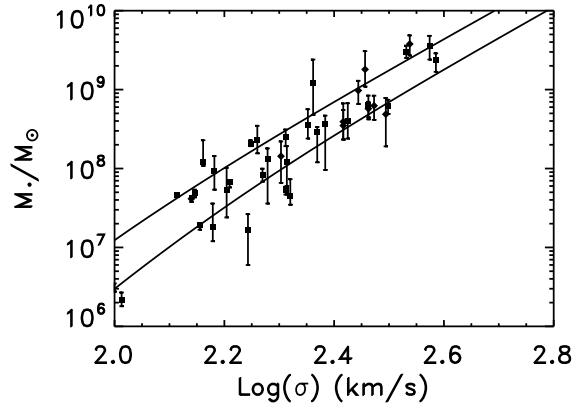,height=6cm,angle=0}}}
\caption{The predicted $M_\bullet$ -- $v_{\rm vir}/\sqrt 2$ relation (lines) compared with the observed
$M_\bullet$ -- $\sigma$ relation (points with error bars), where $\sigma$ is the velocity dispersion of the host bulge. The model is for bremsstrahlung cooling (Eq.~\ref{teq}). The upper line corresponds to a redshift of 
collapse of $z_{\rm c}\simeq 2$. The lower line corresponds to $z_{\rm c}\simeq 0$ and thus to a lower
halo density. The squares with error bars are the black hole mass estimates of \citet{tremaine_etal02}. The diamonds with error bars are those of \citet{ferrarese_merritt00}.
}
\end{figure}

I will illustrate this point by considering the special case of the radio source in M87, at the centre of the Virgo cluster.
If we assume that the Virgo cluster has a mass of $M_{\rm vir}\sim 1.5\times 10^{14}\,M_\odot$ and has virialised at low redshift, then 
$v_{\rm vir}\sim 690{\rm\,km\,s}^{-1}$ and $T_{\rm vir}\sim 10^7\,$K.
If we also assume that the baryons are in the form of a hot gas described by the model of  \citet{komatsu_seljak01}, then the predicted values for the central electron density and the mass of the baryons in the cluster core
are $n_{\rm e}\sim 0.004{\rm\,cm}^{-3}$ and $M_{\rm c}\sim 1.8\times 10^{12}\,M_\odot$ (in reality they will both be lower because some of the baryons are in stars).
The virial temperature $T_{\rm vir}$ is consistent with the X-ray temperature in the core of the Virgo cluster (e.g. \citealp{dimatteo_etal03}), but $v_{\rm vir}/\sqrt 2\sim 500{\rm\,km\,s}^{-1}$ is much larger
than the $\sigma\sim 375{\rm\,km\,s}^{-1}$ determination by \citet{tremaine_etal02} for the stellar velocity dispersion in M87. 
\citet{harms_etal94} and \citet{macchetto_etal97} have estimated that M87 contains a supermassive black hole of  $M_\bullet\sim 3\times 10^9\,M_\odot$.
\citet{dimatteo_etal03} have used this estimate 
for the black hole mass 
together with their X-ray observations of M87 and
have shown that the mechanical power of the jet in M87 can be understood if the black hole accretes the surrounding hot gas at the Bondi
rate ($\alpha\sim 1$) and most of the accretion power is released mechanically ($\beta\sim 1$).  
They have also observed that the central density of the gas in M87 is higher than our estimate 
 from the \citet{komatsu_seljak01} model
 by a factor of $\delta\sim 30-40$.
With these values, Eq.~(\ref{teq}) gives $M_{\rm gas}\sim 2\times 10^{12}\,M_\odot$, which is indeed consistent with our estimate for the mass of the baryons in the core of the Virgo cluster.

With this normalisation issue in mind, the comparison in Fig.~2 demonstrates that this model can reproduce the type of relation observed in the data.
The predicted $M_\bullet$ -- $\sigma$ relation is not exactly a power law, but is consistent with a power law relation of the form $M_\bullet\propto\sigma^\eta$
with $\eta\sim 4$.

\section{Discussion and conclusion}

I have considered a very simple model of self-regulated black hole accretion.
The black hole grows through spherical Bondi accretion and a fraction of the accretion power is 
distributed as thermal feedback into the surrounding IGM.
The heating rate per unit mass of the gas depends on three parameters:
the accretion efficiency with respect to the Bondi rate, $\alpha$, the fraction of the accretion
power converted into heat, $\beta$, and the mass $M_{\rm gas}$ on which the heat is distributed.
If $M_{\rm gas}$ is much larger than the mass within the Bondi radius and
the interaction between the AGN and the IGM happens at large scales (e.g. $\sim 100\,$kpc), 
then one must also deal with the complication that the gas affected by feedback has a different
density from the gas that feeds the black hole. 

If the black hole is not producing enough heat or if the heat is distributed on a very large scale,
the thermal energy input is radiated almost immediately and the equilibrium temperature is not much higher than the IGM temperature without AGN feedback.
For a given metallicity, there is a critical heating rate per unit mass  at which the heating and the cooling rates are comparable
over a broad temperature range. This is the range where the cooling function decreases with the temperature.
Above this critical rate,
the gas is only able to cool at high temperatures, in the bremsstrahlung regime.
For a given set of parameters, the critical heating rate corresponds to a critical black hole mass.
Feedback is inefficient below this critical black mass, while it can rapidly heat the gas to very large temperatures after the black hole has passed this threshold.

In the Bondi model without self-regulation, the black hole grows with $\sim M_\bullet^2$.
Self-regulation reduces the power with which the accretion rate depends on the black hole
mass ($\sim M_\bullet^{0.5}$ for cooling by bremsstrahlung).

If the equilibrium temperature is higher than the
virial temperature, the black hole drives a thermal wind and the approximation that the heating is
quasi-static breaks down.
This blow-out condition implies a maximum $M_\bullet(T_{\rm vir})$, which gives rise to a relation
of the type $M_\bullet\propto v_{\rm vir}^4$.

The question is the relevance of the approximations made by the model:

 i) Bondi spherical accretion in a homogeneous medium supported by thermal pressure
may be a bad estimate of how $\dot{M}_\bullet$ 
depends on the black hole environment, particularly if the gas that feeds the black hole is cold and clumpy.

ii) In a multiphase IGM, the phase which feeds the black hole may not be the same that receives the heat.
One can imagine a scenario in which the AGN is fuelled through the accretion of cold clouds, but most of the AGN heating goes to the dilute hot gas that fills the space between the clouds.
This is the most likely picture in powerful AGN triggered by galaxy mergers.

iii) Feedback may be mechanical and only thermalised on very large scales after 
escaping from the galactic core in a collimated outflow. In our model this corresponds to a large 
value of $M_{\rm gas}$, but the difference between the accretion scale and the
thermalisation scale may determine an oscillation pattern rather than an equilibrium solution
(e.g. \citealp{omma_binney04}).

iv) The onset of a thermally driven wind invalidates the assumption that heating is quasi-static.
In the presence of outflow rather than static boundary conditions, the Bondi formula greatly overestimates 
the black hole accretion rate (e.g. \citealp{dimatteo_etal03}). 
This will almost certainly be the case if the quasi-equilibrium temperature is much higher than the virial
temperature of the system.
 
The toy model presented in this paper is mostly relevant to low power AGNs fed through the accretion of hot gas in the core of massive haloes, but it provides a general rule of thumb to determine
under what conditions AGN feedback is important and self-regulation is possible.

\section{Acknowledgements}

I acknowledge useful conversation with Arman Khalatyan and Romain Teyssier.

\bibliographystyle{mn2e}
\bibliography{references}

\end{document}